\documentstyle[12pt]{article}
\input html.sty
\input epsf
\begin{document}
\title{MATTERS OF GRAVITY, The newsletter of the APS Topical Group on 
Gravitation}

\begin{center}
{\Large {\bf MATTERS OF GRAVITY}}
\bigskip
\hrule
\medskip
{The newsletter of the Topical Group on Gravitation of the American Physical 
Society}
\medskip
{\bf Number 19 \hfill Spring 2002}
\end{center}

\begin{flushleft}

\tableofcontents
\vfill
\section*{\noindent  Editor\hfill}

\medskip
Jorge Pullin\\
\smallskip
Department of Physics and Astronomy\\
Louisiana State University\\
202 Nicholson Hall\\
Baton Rouge, LA 70803-4001\\
Phone/Fax: (225)578-0464\\
Internet: 
\htmladdnormallink{\protect {\tt{pullin@phys.lsu.edu}}}
{mailto:pullin@phys.lsu.edu}\\
WWW: \htmladdnormallink{\protect {\tt{http://www.phys.lsu.edu/faculty/pullin}}}
{http://www.phys.lsu.edu/faculty/pullin}\\
\hfill ISSN: 1527-3431
\begin{rawhtml}
<P>
<BR><HR><P>
\end{rawhtml}

\end{flushleft}
\pagebreak
\section*{Editorial}

Great news are rocking our field of research. Two new NSF-funded
Physics Frontiers Centers, one of them in gravitational wave physics
and another in cosmology. A new privately funded institute devoted to
fundamental physics including gravity is created in Canada. LIGO, TAMA
and GEO are collecting data. A big grid computing project will deliver
computational resources unheard of in the past. Several members of our
community honored.  You can read about it all here, in Matters of
Gravity!

An editorial note: the article by Seiji Kawamura was originally signed
``Seiji Kawamura and the TAMA collaboration'', similarly the article
by Stan Whitcomb was signed ``Stan Whitcomb reporting for the LIGO 
Laboratory and the LIGO Science Collaboration''. I shortened them to
keep in line with usual MOG practice.

Otherwise not much to report here. If you are burning to have Matters of 
Gravity with you all the time, the newsletter is now available for
Palm Pilots, Palm PC's and web-enabled cell phones as an
Avantgo channel. Check out 
\htmladdnormallink{\protect {\tt{http://www.avantgo.com}}}
{http://www.avantgo.com} under technology$\rightarrow$science.
The next newsletter is due February 1st.
If everything goes well this newsletter should be available in the
gr-qc Los Alamos archives 
(\htmladdnormallink{{\tt http://xxx.lanl.gov}}{http://xxx.lanl.gov})
under number gr-qc/0202044. To retrieve it
send email to
\htmladdnormallink{gr-qc@xxx.lanl.gov}{mailto:gr-qc@xxx.lanl.gov}
with Subject: get 0202044
(numbers 2-17 are also available in gr-qc). All issues are available in the
WWW:\\\htmladdnormallink{\protect {\tt{http://www.phys.lsu.edu/mog}}}
{http://www.phys.lsu.edu/mog}\\ 
A hardcopy of the newsletter is
distributed free of charge to the members of the APS
Topical Group on Gravitation upon request (the default distribution form is
via the web) to the secretary of the Topical Group. 
It is considered a lack of etiquette to
ask me to mail you hard copies of the newsletter unless you have
exhausted all your resources to get your copy otherwise.
\par
If you have comments/questions/complaints about the newsletter email
me. Have fun.
\bigbreak

\hfill Jorge Pullin\vspace{-0.8cm}
\section*{\centerline {
We hear that...}}
\addcontentsline{toc}{subsubsection}{\it  
We hear that... by Jorge Pullin}
\begin{center}
Jorge Pullin
\htmladdnormallink{pullin@phys.lsu.edu}
{mailto:pullin@phys.lsu.edu}
\end{center}

{\em J\"urgen Ehlers} will receive the 2002 Max Planck Medal of the German
Physical Society. The medal is given by the Society for outstanding
acheivements in theoretical physics from world-wide nominations and is
widely regarded as its the highest international honor. It was
instituted in Jnue 1929 to mark the 50th anniversary of Planck's
doctorate. The first two recipients were Planck and Einstein. Ehlers
will receive the prize in March 2002 during a special ceremony at the
annual meeting of DFG.

{\em Gary Horowitz, Bei-Lok Hu} and {\em David Shoemaker} have been 
made Fellows of the American Physical Society.

Hearty congratulations!

\vfill
\pagebreak

\section*{Correspondents}
\begin{itemize}
\item John Friedman and Kip Thorne: Relativistic Astrophysics,
\item Raymond Laflamme: Quantum Cosmology and Related Topics
\item Gary Horowitz: Interface with Mathematical High Energy Physics and
String Theory
\item Beverly Berger: News from NSF
\item Richard Matzner: Numerical Relativity
\item Abhay Ashtekar and Ted Newman: Mathematical Relativity
\item Bernie Schutz: News From Europe
\item Lee Smolin: Quantum Gravity
\item Cliff Will: Confrontation of Theory with Experiment
\item Peter Bender: Space Experiments
\item Riley Newman: Laboratory Experiments
\item Warren Johnson: Resonant Mass Gravitational Wave Detectors
\item Stan Whitcomb: LIGO Project
\item Peter Saulson: correspondent at large.
\end{itemize}
\vfill
\pagebreak

\section*{\centerline {
Center for Gravitational Wave Physics,}\\
\centerline{a new NSF Physics Frontier Center}}
\addtocontents{toc}{\protect\medskip}
\addtocontents{toc}{\bf Community news:}
\addtocontents{toc}{\protect\medskip}
\addcontentsline{toc}{subsubsection}{\it  
Center for Gravitational Wave Physics, by Sam Finn}
\begin{center}
Lee Samuel Finn, PennState
\htmladdnormallink{lsfinn@psu.edu}
{mailto:lsfinn@psu.edu}
\end{center}

On August 15 the National Science Foundation created at Penn State the
Center for Gravitational Wave Physics as part of its new Physics
Frontier Centers program.

As described by the Foundation, the goal of the Physics Frontier Center
(PFC) program is to support timely, aggressive, and forward-looking
research with the potential to lead to fundamental advances in
physics. This new Center is one of only three funded by the Foundation
in the first round of Physics Frontier Center funding.  The mission of
the Center for Gravitational Wave Physics is to help crystallize and
develop the emerging discipline of gravitational wave phenomenology:
the astrophysicsÊ and fundamental physics that gravitational wave
observations --- in all wavebands --- enable.

Research at the Center will focus on interdisciplinary problems
at the interface of general relativity, gravitational waves, astrophysics
and detector design:
\begin{itemize}
\item Astrophysics and gravitational waves:
problems of source calculations, astrophysical modeling of sources and
their populations, and interpretation of observations,

\item General relativity and gravitational waves: testing relativity
and developing numerical and analytic tools needed for detailed
studies of sources (e.g., numerical relativityÊ and radiation
reaction),

\item Detector design studies: how target science --- the sourcesÊ
one wants to detect or the science one wants to do --- constructively
influences the design of advanced gravitational wave detectors.
\end{itemize}

\textit{The Center for Gravitational Wave Physics is a community
resource,} meant to support and encourage research in gravitational
wave phenomenology. An important component of the Center is a major,
international visitors program. Center funding is available to host
visitors or groups of visitors who wish to become involved in
gravitational wave phenomenology research or focus attention on
specific problems. Visits, supported by the Center, from weeks to
months are possible, and limited funding is available to support
sabbatic visitors.

In addition, the Center will host frequent focus sessions,
workshops and conferences on critical gravitational wave phenomenology
problems.  Focus sessions, which last for just a few days, typically
address a single, narrowly defined problem on which it is thought
substantial progress can be made through concentrated effort by experts. 
Workshops, like the recent Gravitational Wave Phenomenology Workshop
(described elsewhere in this volume), last from a few days to a week are
broader meetings, aimed at discussion and accessible to non-experts and
new-comers to the field.  Conferences, such as the forthcoming Fourth
International LISA Symposium, are larger and longer affairs, whose purpose
is to consolidate work in preparation for future efforts. 

The recent Gravitational Wave Phenomenology Workshop, held on 6--8
November 2001, was the first workshop sponsored by the new
Center. Forthcoming focus sessions include
\begin{itemize}
\item Astrophysical Initial Data Problem: (29--30 March 2002) Posing
Astrophysically relevant initial data for numerical relativity
investigations of binary black hole or neutron star
coalescence. Organized by Greg Cook and Pablo Laguna.

\item WORK-BENCH: (Spring 2002) Present use and future development of
the \texttt{bench} program for advanced interferometric detector
design. Organized by Sam Finn, Gabriela Gonz\'alez, David Shoemaker,
Robin Stebbins and Ken Strain.

\item Radiation Reaction: (Spring 2002) Implementing practical schemes
for computing the gravitational waveforms, especially from extreme
mass ratio binary systems. Organized by Warren Anderson, Patrick
Brady, Eanna Flanagan and Sam Finn.

\item Numerical Relativity: (24--29 June 2002) Jointly sponsored with
the Institute for Mathematics and its Applications (IMA), this
workshop will bring together the numerical relativity and the
mathematicians working in numerical analysis, scientific computation,
partial differential equations and geometry for an intense but
informal discussion aimed at bringing new ideas and techniques into
the numerical relativity, and propelling applied mathematicians with
relevant skills and interest into numerical relativity. Organized by
IMA director Doug Arnold, Abhay Ashtekar and Pablo Laguna.

\item Stellar Populations: (Fall 2002) What --- and how --- can we
learn about stellar populations from gravitational wave observations?
Organized by Vicky Kalogera, Martin Rees and Sam Finn.

\item Massive black hole coalescence: (Fall 2002) Massive black holes
are presumed to coalesce in the cores of interacting galaxies, and
these coalescence events are potentially important gravitational wave
sources for LISA. Present theoretical estimates of the coalescence
rates give timescales much longer than suggested by indirect
observational evidence. What's missing from our understanding?
Organized by Steinn Sigurdsson and Ramesh Narayan.

\item Numerical Relativity and Gravitational Wave Data Analysis: (Fall
2002) Numerical relativity has much to offer to the analysis and
interpretation of gravitational wave observations. This focus session
will bring these two communities together to foster a greater
understanding of how numerical relativity can aid in gravitational
wave data analysis and interpretation. Organized by Bernd Br\"ugmann,
Sam Finn and Pablo Laguna.

\end{itemize}
Forthcoming workshops hosted by the Center include the Fourth Capra
Meeting on Radiation Reaction in General Relativity, which will follow
immediately on the heels of the radiation reaction focus session, and
the second Gravitational Wave Phenomenology Workshop, tentatively
planned for Spring 2003. Forthcoming conferences hosted by the Center
include the Fourth International LISA Symposium, which will be held
19--24 July 2002.

The core, resident faculty of the Center for Gravitational Wave Physics are
Abhay Ashtekar, Sam Finn (Director), Peter Meszaros, Pablo Laguna
(Associate Director), Steinn Sigurdsson and Alex Wolszczan.  In addition,
the Center for Gravitational Wave Physics non-resident facultyÊ members,
who are expected to visit frequently, are Warren Anderson, Mario Diaz and
Joseph Romano (University of Texas, Brownsville); Patrick Brady (University
of Wisconsin, Milwaukee); Matt Choptuik (University of British Columbia);
Eanna Flanagan (Cornell University); Gabriela Gonzalez, Jorge Pullin and
Joel Tohline (Louisiana State University); Richard Price (University of
Utah); Robin Stebbins (Goddard Spaceflight Center); and Ken Strain
(University of Glasgow).

All Center activities are open to the broad scientific community,
whose participation will be supported through the Center's visitor
program.  For more information on the opportunities provided by the
Center please contact \texttt{CGWP@Gravity.Phys.PSU.Edu} or see the
Center's web site (presently under construction) at
\htmladdnormallink{{\tt http://cgwp.phys.psu.edu}}{http://cgwp.phys.psu.edu}

\vfill
\pagebreak
\section*{\centerline {
Perimeter Institute for Theoretical Physics}}
\addcontentsline{toc}{subsubsection}{\it  
Perimeter Institute for Theoretical Physics, by Lee Smolin}
\begin{center}
Lee Smolin, Perimeter Institute and University of Waterloo
\htmladdnormallink{lsmolin@perimeterinstitute.ca}
{mailto:lsmolin@perimeterinstitute.ca}
\end{center}

Perimeter Institute for Theoretical Physics opened its doors this past
September in Waterloo, Ontario, Canada.  Pi, as we call it, is an
independent, privately endowed institute, which will specialize in
fields having to do with fundamental physics.  This naturally includes
several fields in which relativists work including quantum gravity,
cosmology and string theory.

This year Pi opened with 3 researchers (long term positions equivalent
to faculty), two associates, who have appointments also in a new
Institute of Quantum Computing (IQC) founded simultaneously at the
University of Waterloo, four postdocs and 5 students. (For the names
and much other information please see 
\htmladdnormallink{{\tt http://www.perimeterinstitute.ca}}
{http://www.perimeterinstitute.ca}). The
scientific staff is being recruited primarily from outside Canada.
Next year we expect to have around 7 researchers, 10 postdocs, plus
visitors and students. There is also an active visitors and seminar
programs, with an average of 2 new visitors arriving each week (for
details see 
\htmladdnormallink{{\tt 
http://www.perimeterinstitute.ca/news\_{}fr.htm}}
{http://www.perimeterinstitute.ca/news\_{}fr.htm}).

There is already quite a lively atmosphere, and the visitor to Pi will
usually find several informal discussions and collaborations under way
throughout the building.  In January alone we had seminars by Stephon
Alexander (London), Giovanni Amelino-Camelia (Rome), Daniel Gottesman
(Berkeley), Ted Jacobson (Maryland), Robert Laughlin (Stanford), Seth
Major (Hamilton), Hendryk Pfieffer (Cambridge) and Maxim Pospelov
(Victoria), who presented their work on a range of topics in quantum
gravity, quantum phase transitions, black hole physics, Planck scale
phenomenology and quantum information theory.

The plan is to grow over six years to 40 resident scientists, plus
visitors, students, associates and affiliates (the latter are people
whose primary appointments are in nearby universities.) The fields of
emphasis for this first stage are quantum gravity, including but not
limited to string theory and foundations of quantum mechanics
including quantum information theory. In later years other fields of
theoretical physics will be added.

A beautiful new building is planned, designed by the architectural
firm of Saucier and Perrott, who were chosen after a competition. The
building was designed after extensive consultations with scientists
and we believe will provide the most hospitable and welcoming
atmosphere that exists for doing theoretical physics, in a visually
stunning setting (see
\htmladdnormallink{{\tt 
http://www.perimeterinstitute.ca/news\_{}fr.htm}}
{http://www.perimeterinstitute.ca/news\_{}fr.htm}).
). The building will be on
the side of Silver Lake, next to Waterloo Park, and a short walk from
both the University of Waterloo and the newly invigorated downtown
core of the city of Waterloo. While Pi will be primarily a residential
institute, there will be conferences, workshops and the like held in
the building, as well as cultural activities such as concerts and
lectures on science for the public. The first Pi public lecture was
given in October by Roger Penrose, and drew an audience that was twice
the capacity of the lecture hall.

Presently we are located in a beautiful old red stone building in the
center of Waterloo.  The building features a large informal
interaction area with, we believe, the only bar in the world with
wall-to-wall black boards (called by popular acclaim the hbar).

Waterloo is in the heart of the fastest growing region in Canada, and
is a center for high tech industry as well as home to two
universities. Those of us who have just moved here have been very
pleased to find ourselves in a sophisticated, youthful, growing and
diverse city.  Toronto, which is one of the most exciting cities
culturally in North America, and is also the world's most
ethnically diverse city, is a bit more than an hour away by car, bus
and train. The Toronto International airport is 45-50 minutes away.

The Institute was founded by Mike Lazaridis, founder and co-Ceo of
Research in Motion and several friends, who are contributing an
endowment of \$120 million (Canadian).  The Institute is governed
primarily by its scientific staff, with oversight from a Scientific
Advisory Committee (SAC) of renowned senior theoretical physicists.
As in private universities, there is a board of directors, but
scientific decisions are primarily the purview of the scientific staff
with oversight by the SAC.  While independent, the institute has
negotiated already or has under discussion a complex set of
relationships with the University of Waterloo and other universities
in Southern Ontario and Canada. These relationships include cross
appointment of researchers, associates, affiliates and joint
projects. The internal scientific governance is democratic and
non-hierarchical; there is no scientific director and there are no
heads of groups. While this condition was mandated by the founders, it
was also discussed extensively among the scientists and advisory
committee, and a number of models and examples were studied before
arriving at the present structure. We believe that history shows that
the scientific institutes and departments that maintain themselves at
the highest level of quality for the longest time are those run on
relatively non-hierarchical and democratic lines.

But just as important as money, bricks, offices, networks and
computers is spirit, philosophy and culture.  We are designing Pi with
the hope that it will remain perpetually youthful, dynamic and
flexible, a home to important research on the frontiers of physics,
even when it is no longer young. Pi scientists are chosen to be not
only scientific leaders in their fields, but dynamic, risk taking,
open and ambitious people, who are interested in work done outside
their specializations and are open to new ideas and competing research
programs.  They must also be people who respect other people,
communicate easily and honestly, prefer working in a non-hierarchical
and democratic setting and are interested in being involved in the
adventure of building a new scientific institute.

Of course, in the end, the only measure of success of a new institute
is the quality of the science that is done there, sustained over many
years. At present, all that can be said is that Pi is off to a very
good start, and we hope to see and host many members of the
gravitational physics community here over the next few years. Watch
this space.

For more information regarding seminars and visitors and information
about postdoctoral, visiting and long term positions see
\htmladdnormallink{{\tt 
http://www.perimeterinstitute.ca/news\_{}fr.htm}}
{http://www.perimeterinstitute.ca/news\_{}fr.htm}).

\vfill
\pagebreak
\section*{\centerline {
Detector and Data Developments within GEO 600}}
\addtocontents{toc}{\protect\medskip}
\addtocontents{toc}{\bf Research Briefs:}
\addtocontents{toc}{\protect\medskip}
\addcontentsline{toc}{subsubsection}{\it  
Detector and Data Developments within GEO 600, by Alicia Sintes}
\begin{center}
Alicia M. Sintes, Albert Einstein Institut
\htmladdnormallink{sintes@aei-potsdam.mpg.de}
{mailto:sintes@aei-potsdam.mpg.de}
\end{center}

  It is not the landscape that makes Ruthe, near Hannover, Germany, an 
exciting place. The site of the British-German GEO-600  gravitational wave 
detector is a peaceful and relaxed placed. Unlike Hanford and Livingston,
which host its larger LIGO brethren, there is no danger of earthquakes 
or wildfires, radioactivity is low, and alligators have not been seen 
around for a very long time.

  GEO is the first medium-scale interferometer to enter in operation in 
Europe. By mid 2002, GEO plans to achieve its full sensitivity. Although 
GEO's armlength is only 600 meters, compared to LIGO 4 km, it features 
advanced mirror suspensions and optics which, by the way, are planned to be 
incorporated into advanced LIGO. The use of such advanced technology makes 
GEO design sensitivity almost comparable to initial LIGO, and therefore it 
has reasonable chances of detecting gravitational waves (especially from 
compact binary inspirals, neutron stars and supernova events). Moreover, 
GEO will be unique among the first interferometers in being able to operate 
in narrow-band mode which will give it a better sensitivity to continuous, 
nearly periodic signals than the larger projects (LIGO and the French-Italian
VIRGO) would have in the selected band.

  While getting closer to completion, 2001 has been an intensive year for 
GEO. Detector and environmental data have been produced continuously
throughout the year. GEO generates on the order of 50 GBytes of data per 
day, which are stored using a data format fully compatible  with LIGO and 
VIRGO. In addition, GEO and LIGO have become partners: they have signed 
full reciprocal data exchange agreements, GEO is part of the LSC (LIGO 
Scientific Collaboration), and they are jointly developing data analysis 
software. Besides, GEO is building four different computer clusters at 
AEI-Golm, Birmingham, Cardiff and Hannover where the different data 
analysis tasks are to be performed.

  In February 2001, a workshop took place in Ruthe and gave the main kick to 
GEO Detector Characterization (DC) activities. Those who attended realized 
that the site is cold in winter, especially if you need to walk to the next
building equipped with facilities, which highlights the enthusiasm 
experimentalists are putting in bringing the interferometer to work. Later on, 
in June, another GEO-DC workshop took place, but this time at AEI-Golm. On 
that occasion, the need of the DC-Robot, as an efficient automated system to 
characterize the data, and its interaction with a database were pushed forward.

  Another effort, largely invisible to outsiders, is the analysis of GEO 
data. With the aim of gaining a better understanding of the detector 
behavior and its environment, data from environmental monitors, e.g., 
seismometers, magnetometers, and, of course, from the detector itself are
being analyzed. As the the detector status has been progressing, data 
analysis activities have turned more complex and organized. After the 
summer, different subgroups focused their attention to the commissioning of 
different detector subsystems. Of particular interest was the GEO engineering
run that took place in October 15-18, 2001. The mode-cleaners were in almost
final configuration and the Michelson was locked on mid fringe with no power
recycling. This run was a success, shift scheme and data transfer 
Ruthe-Hannover-AEI were exercised, and the long term behavior of the whole 
system was tested.

  As in any other project, GEO has overcome many hardships, some of them 
unexpected. At the end of October, data acquisition at the north-end station
was interrupted for several days because mice had eaten too many optical 
fibers. This time, new fibers with a 40 year guarantee of mouse-proofness 
were ordered to give us the upper hand in the fight.
 
  From November onwards, a big effort was devoted to lock the Michelson
on a dark fringe and incorporating the power recycling cavity. In parallel, 
a deep analysis of the detector subsystems (geophones, laser power, 
magnetometers, mode-cleaners, etc) has taken place. To keep track of the 
detector and data acquisition status, GEO detector database is available via 
a web interphase. This includes also signal descriptions, calibration 
information, data viewer, and two electronic labbooks (GEO-600 and 
GEO-DC) actively used by both experimentalists and theorists.

  It is worth mention, although not being a gravitational wave detection, 
GEO first astronomical observation has already occurred. The geomagnetic 
storm due to two fast moving coronal mass ejections on November 22, 2001, 
was observed in the GEO magnetometers as expected in November 24 data.

  The last thing I want to mention and most exciting one, is the coincidence
run with LIGO. Both LIGO and GEO detectors were on operation from December 
28th until January 14th. The run was a great success. GEO had a duty 
cycle of about 80\%. For some days the power-recycled interferometer was
in lock for more than 95\% of the time. The longest continuous
lock segment was of 3h 48min.  Monday 14th was a day for celebration in
Ruthe with all the operators and people that participated in the shifts.
We really got very good data!

 During the run other events took place. For example, on January 2nd, the 
waves from an earthquake near Australia (Vanuatu Islands, magnitude 7.1 on 
the Richter scale) hit GEO. The interferometer lost lock, but it was 
realigned and locked automatically from then on. Almost all the earthquakes 
worldwide, with a magnitude bigger than 4.5, can be seen on our
seismometer data but often do not influence the detector output so much.
In the data (e.g., from the feedback to the intermediate masses) we can 
clearly see the influence of the moon with a period of 12.4h, and our 
magnetometers continue recording information of the Sun-Earth environment. 
With the gravitational wave data taken from this coincidence run we hope to 
be able to set astrophysical upper limits on different gravitational wave 
sources. 

  Pay attention to the next Matters of Gravity edition, I am sure there will 
be plenty of news from the gravitational wave community.

\vfill
\pagebreak
\section*{\centerline {Grid Physics, the Virtual Data Grid, and LIGO}}
\addcontentsline{toc}{subsubsection}
{\it The Virtual Data Grid and LIGO, by Pat Brady and Manuela
Campanelli}
\begin{center}
Patrick Brady (University of Wisconsin--Milwaukee) and \\
Manuela Campanelli (The University of Texas at Brownsville)
\htmladdnormallink{patrick@gravity.phys.uwm.edu}
{mailto:patrick@gravity.phys.uwm.edu}
\htmladdnormallink{manuela@aei-potsdam.mpg.de}
{mailto:manuela@gluon.utb.edu}
\end{center}

Between 28 December 2001 and 14 January 2002, the three largest
interferometric gravitational-wave detectors in the world were
listening for signatures of cataclysmic astrophysical events in
our Galaxy and beyond.    For many involved in the LIGO (Laser
Interferometer Gravitational-wave Observatory) project,  this was an
event of grand proportions which demonstrated that we are truly on
the brink of gravitational-wave astronomy.   Yet the data run was just
the beginning.   About 13 Terabytes of data was recorded and will be
analyzed over the coming months.    Scaling these numbers to full
scale scientific operations,  the experiment will generate several
hundreds of Terabytes of data per year.   

The variety of sources which could produce gravitational waves call
for a variety of search techniques to be applied to the data stream.
For example,  searches for stochastic background gravitational waves
require minimal computational power--a standard desktop workstation is
good enough--yet,  the search must access the data from all the
detectors.   At the other end of the spectrum of computational
requirements are searches for continuous signals from spinning neutron
stars.  These signals are extremely weak,  and require coherent
accumulation of signal-to-noise for long periods of time;   this
introduces the need to account for earth-motion-induced Doppler shifts
and internal evolution of the sources.  Thus,  the variety of signals
and their weakness lead to an analysis problem which can use
essentially infinite computing resources.   

But the computational
requirements are only half the story.  LIGO computing facilities and
scientific users reside at many different national and international
centers and universities.  For the LIGO Scientific Collaboration (LSC)
[1], therefore,  accessing these large datasets and performing
an efficient analysis on them requires a dynamically distributed
computational infrastructure, including tools to manage storage,
migration and replication of data, job control, and cataloging of the
many data products.   The LIGO Data Analysis System handles these
problems on a scale consistent with the LIGO-I mission,  however
\emph{Grid Computing} [2] provides a new computational
infrastructure to extend and enhance current capabilities to a level
consistent with the expected requirements. 

LIGO is only one of several physics experiments expecting to generate
vast amounts of data which must be carefully analyzed using complex
algorithms requiring enormous computational power.      For this
reason several LSC member institutions, including California
Institute of Technology (CIT), The University of Texas at Brownsville
(UTB), and University of Wisconsin--Milwaukee (UWM), are
participating in a multi-experiment project, sponsored by the
National Science Foundation,  to build the first Petabyte-scale
computational grid environment for data intensive experiments.  The
Grid Physics Network (GriPhyN) [9] project is a collaboration of both
experimental physicists and information technology researchers.
Driving the project are unprecedented requirements for
geographically dispersed extraction of complex scientific information
from very large collections of measured data, flowing from four
experiments in high-energy and nuclear physics (two large hadron
colliders at CERN, CMS and ATLAS [3]), gravitational waves
(LIGO [4]) and astronomy (the SDSS project [5]).  To
meet these requirements, GriPhyN researchers will develop the Virtual
Data Toolkit (VDT) containing basic elements to construct the first
Global Petascale Virtual Grid.  

The virtual data concept aims to unify the view of data in the
distributed Grid environment.  It will not matter if the data is raw
or processed, or  if it was generated from an hadron collider experiment
or a gravitational-wave detector. The virtual Grid will enable data
access and archival at nodes distributed around the globe while storing
meta-data which make the data self-describing.    The VDT developed by
GriPhyN will be deployed on the Grid to directly manage these
fundamental virtual data objects instead of complex data pipelines.
In the case of LIGO, the VDT will be capable of executing deep
searches for gravitational waves using many machines distributed
around the world,  while making the results available to the
scientists in a transparent fashion.   Once deployed,  the Grid tools
currently under development will significantly enhance scientists'
ability to carry out the necessary analysis of LIGO data.  In fact,
prototype data replication tools (being developed by Scott Koranda at
UWM) are already moving data from the archive at Caltech onto spinning
disks at UWM for analysis using the UWM system.  

GriPhyN is not the only data grid project, although it is one of the
largest and probably most advanced in the world.   Similar projects
are now active in Europe and Asia.  In September, the NSF announced
the additional award of \$13.65M over five years to a consortium of 15
universities and four national laboratories to create the
International Virtual Data Grid Laboratory [6,7] (iVDGL).
The iVDGL, to be constructed in partnership with the European Union,
Japan, Australia and eventually other world regions, will form the world's
first true {\it Global Grid}. The iVDGL will provide a unified computational
resource for major scientific experiments in physics, astronomy,
biology, and engineering.  The iVDGL will therefore serve as a unique
computing resource for testing new GriPhyN
computational paradigms at the Petabyte scale and beyond.  Management
of the iVDGL is integrated with that of the GriPhyN project.  The
international partners are investing more than \$20M around the world
to build computational sites as part of the consortium.
Moreover, the NSF award of iVDGL is matched by \$2M in university
contributions, plus funding for Computer Science Fellows by the UK
e-Science Programme [10].  Of this total award,  \$2.11M
will go to universities affiliated with the LIGO Laboratory to develop
Grid Computing centers at the three GriPhyN/LSC institutions (CIT, UWM
and UTB) and at Pennsylvania State University (PSU).  

A significant challenge for science in the 21st century is data
management and analysis.  Just as large database technology has
revolutionized the commercial world as the backbone of many
information intensive enterprises,  so virtual data,  Grid computing
and transparent access to a world of computing resources will
revolutionize science in the coming decade. 

\bigskip

{\bf References:}

[1] LIGO Scientific Collaboration web site: \\
\htmladdnormallink{{\tt 
http://www.ligo.caltech.edu/LIGO\_{}web/lsc/lsc.html}}
{http://www.ligo.caltech.edu/LIGO\_{}web/lsc/lsc.html}.

[2] The Computational Grid is described in the book 
``The Grid : Blueprint for a New Computing Infrastructure" edited by 
Ian Foster 
and Carl Kesselman -- Morgan Kaufmann Publishers (1998) ISBN 1-55860-475-8.
Many more references can be found at the following web site:\\
\htmladdnormallink{{\tt 
http://www.aei-potsdam.mpg.de/\~{}manuela/GridWeb/info/grid.html}}
{http://www.aei-potsdam.mpg.de/\~{}manuela/GridWeb/info/grid.html}.

[3] CMS and ATLAS are two large hadron colliders at CERN, the world's 
largest particle physics center near Geneva in Switzerland (see web site:\\
\htmladdnormallink{{\tt 
http://www.griphyn.org/info/physics/high.html}}
{http://www.griphyn.org/info/physics/high.html}
.)

[4] The Laser Interferometer Gravitational-wave Observatory 
web site: \\
\htmladdnormallink{{\tt 
http://www.ligo.caltech.edu}}
{http://www.ligo.caltech.edu}.

[5] The Sloan Digital Sky Survey (SDSS) project  
is the most ambitious astronomical
survey project ever undertaken. The survey will map in detail
one-quarter of the entire sky, determining the positions and absolute
brightnesses of more than 100 million celestial objects. It will also
measure the distances to more than a million galaxies and
quasars. Apache Point Observatory, site of the SDSS telescopes, is
operated by the Astrophysical Research Consortium (ARC).
(see web site at:
\htmladdnormallink{{\tt 
http://www.sdss.org/sdss.html}}
{http://www.sdss.org/sdss.html}
)

[6] The International Virtual Data Grid Laboratory web site:\\ 
\htmladdnormallink{{\tt http://www.ivdgl.org}}
{http://www.ivdgl.org}.

[7] The Outreach Center of the Grid Physics Network web site:\\ 
\htmladdnormallink{{\tt 
http://www.aei-potsdam.mpg.de/\~{}manuela/GridWeb/main.html}}
{http://www.aei-potsdam.mpg.de/\~{}manuela/GridWeb/main.html}.

[8] The Grid Physics Network web site: 
\htmladdnormallink{{\tt 
http://www.griphyn.org}}{http://www.griphyn.org}.

[9]  e-Science is an equivalent project to iVDGL in the UK (see web site at:\\
\htmladdnormallink{{\tt 
http://www.e-science.clrc.ac.uk/}}
{http://www.e-science.clrc.ac.uk/}).

\vfill
\pagebreak
\section*{\centerline {
1000 hours of data taken on TAMA300}\\
\centerline{and the first lock of
the recycled TAMA300}}
\addcontentsline{toc}{subsubsection}{\it  
1000 hours of data and first lock of the recycled TAMA300, by Seiji Kawamura}
\begin{center}
Seiji Kawamura, National Astronomical Observatory of Japan 
\htmladdnormallink{seiji.kawamura@nao.ac.jp}
{mailto:seiji.kawamura@nao.ac.jp}
\end{center}

The TAMA project, the Japanese effort for detecting gravitational
waves using the 300m laser interferometer (TAMA300), took an
unprecedented 1000 hours of data in the summer of 2001. Just
recently the power recycling system was implemented in TAMA300,
and the recycled interferometer was successfully locked.

During the observation period, the interferometer was remarkably
stable: it held lock continuously for more than 20 hours several
times, and the overall duty cycle was 86\%. The observational
functions of the detector had been drastically improved for this
data run: a newly-developed automatic re-locking system of the
whole interferometer worked reliably, a newly-established quick
look system helped us to find any unusual behavior of the
interferometer as well as the data taking system, and a
newly-implemented medium-speed data acquisition system (64
channels) supplemented the existing high-speed and low-speed data
acquisition system (100 channels) for recording important
detector information. As for the sensitivity of the detector, it
had been improved around 100 Hz by a factor of 10 compared with
the sensitivity obtained in the summer of 2000, resulting in a
significant improvement of the sensitivity to chirps from
heavier-mass binary coalescence. The best strain sensitivity of
$5x10^{-21} {\rm Hz}^{-1/2}$ around 
1 kHz remained the same as before.

During the above-mentioned data run, the interferometer was
operated without the power recycling system. Since then we have
begun implementing recycling in TAMA300. Around the end of 2001
the recycled interferometer was finally locked for a few seconds
for the first time in the history of TAMA300! The lock has been
made more and more robust by re-activating the alignment control
system for the test masses and by adjusting all the servo systems
carefully. As of Jan. 23, 2002, TAMA300 with recycling can hold
lock for up to 46 minutes continuously. We will continue to
stabilize the lock of the interferometer as well as to improve
the sensitivity of the detector.

Please have a look at our home page for more details,
\htmladdnormallink{{\tt 
http://tamago.mtk.nao.ac.jp}}{http://tamago.mtk.nao.ac.jp}

\vfill
\pagebreak
\section*{\centerline {LIGO Takes Some Data!}}
\addcontentsline{toc}{subsubsection}{\it  
LIGO Takes Some Data!, by Stan Whitcomb}
\begin{center}
Stan Whitcomb
\htmladdnormallink{stan@ligo.caltech.edu}
{mailto:stan@ligo.caltech.edu}
\end{center}

Many members of the LIGO Scientific Collaboration (LSC) spent
The final days of 2001 and the early days of 2002 at the LIGO 
observatories in Hanford,
Washington, and Livingston, Louisiana, participating in the
seventh LIGO engineering run (E7). Unlike previous LIGO
engineering runs which focused on characterizing the
interferometers and improving their reliability, the goal of E7
was to provide data for a first end-to-end test of the data
analysis pipeline, to test data acquisition and archiving, gain
experience with round-the-clock interferometer operation and 
detector monitoring.

For two weeks teams of scientists and operators attempted to keep 
the three LIGO interferometers locked and taking data. Although the 
LIGO interferometers still have a long way to go to reach their design
sensitivity, the data recorded during E7 will provide LSC
scientists with the real interferometer data needed for perfecting
and tuning their gravitational wave search algorithms.
In addition to the three LIGO
interferometers, we were fortunate to have the GEO-600
interferometer near Hanover, Germany and the Allegro bar-detector 
at Louisiana State University operating in cooperation with LIGO
during much of the run.

To maximize the overlap of the lock times among the interferometers
each LIGO interferometer was operated in a configuration that minimized
the risk of down time. The 2~km interferometer at Hanford was
operated in its final power-recycled mode. Because the
commissioning of the 4~km interferometers at
Hanford and Livingston has been scheduled to lag that of the 2~km
instrument, their power-recycling mode is not yet reliable enough
for extended data taking. The 4~km instruments were therefore
operated in a non-recycled mode to improve overall lock-time. 

The E7 run, like previous runs, was an occasion for large number
of LSC scientists to participate actively in the operation of the
interferometers and to perform other scientific activities at the
observatories. 
Monitoring programs ran continuously to help the operators and
scientists keeping tabs on the current interferometer status, and
preliminary analysis of the data collected during the E7 run took
place using the LIGO Data Analysis Systems (LDAS) at the sites. In
addition, off site analysis occurred on smaller selected data sets
at Caltech and MIT. In the end, over 13 TB of data from nearly 8000 
interferometer and environmental channels had 
been collected and archived.

The three interferometers were individually locked for 
$\sim$60-70\% of the run, with the majority of that time 
($\sim$40-60\% of the total) in locked segments long enough
for meaningful analysis (more than 15 minutes).  The total time
with all three interferometers locked was 140 hours ($\sim$34\%), 
out of which 71 hours ($\sim$18\%) represent segments longer 
than 15 minutes.
Considering the early stage in the commissioning, we are rather 
pleased with this performance.

LIGO is funded by the National Science Foundation under
Cooperative Agreement PHY--9210038. This work is a collaborative
effort of the Laser Interferometer Gravitational-wave Observatory
and the institutions of the LIGO Science Collaboration.

More information about LIGO can be found at: 
\htmladdnormallink{{\tt 
http://www.ligo.caltech.edu}}{http://www.ligo.caltech.edu}.

{\centerline {LIGO-T020016-00-D} }

\vfill
\pagebreak
\section*{\centerline {Quantum gravity: progress from an unexpected
direction}}
\addcontentsline{toc}{subsubsection}{\it  
Quantum gravity: progress from an unexpected direction, by Matt Visser}
\begin{center}
Matt Visser, Washington University 
\htmladdnormallink{visser@wuphys.wustl.edu}
{mailto:visser@wuphys.wustl.edu}
\end{center}

Over the last few of years, a new candidate theory of quantum gravity
has been emerging: the so-called ``Lorentzian lattice quantum
gravity'' championed by Jan Ambjorn [Niels Bohr Institute], Renate
Loll [Utrecht], and co-workers [1]. It's not brane
theory (string theory), it's not quantum geometry (new variables); and
it's not traditional Euclidean lattice gravity. It has elements of
both the quantum and the geometric approaches; and it is sufficiently
different to irritate partisans of both camps.

Quantum gravity, the as yet unconsummated marriage between quantum
physics and Einstein's general relativity, is widely (though perhaps
not universally) regarded as the single most pressing problem facing
theoretical physics at the turn of the millennium. The two main
contenders, ``Brane theory/ String theory'' and ``Quantum geometry/
new variables'', have their genesis in different communities. They
address different questions, using different strategies, and have
different strengths (and weaknesses).

Brane theory/ string theory grew out of the high-energy particle
physics community, and views quantum physics as
paramount [2]. The consensus feeling in the brane
community is that to achieve the quantization of gravity they would be
willing to take quite drastic steps, to mutilate the geometrical
foundations of general relativity and if necessary to {\emph{force}}
general relativity to fit into the brane framework. In contrast, the
general relativity community views the geometrical nature of
Einstein's gravity as sacrosanct, and would by and large be quite
willing to do a little drastic surgery to the foundations of quantum
physics if they felt it unavoidable [2]. ``Lorentzian
lattice quantum gravity'' does a little of both: it adopts some
aspects of each of these approaches, and violates other cherished
notions of these two main candidate models.

On the one hand, ``Lorentzian lattice quantum gravity'' has grown out
of the lattice community, itself a subset of the particle physics
community. In lattice physics spacetime is approximated by a discrete
lattice of points spaced a finite distance apart. This
``latticization'' process is a way of guaranteeing that quantum field
theory can be defined in a \emph{finite} and \emph{non-perturbative}
fashion. (Indeed currently the lattice is the {\emph{only}} known
non-perturbative regulator for flat-space quantum field theory. This
technique is absolutely essential when carrying out computer
simulations of quantum field theories, and in particular, computer
simulations of quarks, gluons, and the like in QCD.)  In addition to
these particle physics notions, ``Lorentzian lattice quantum gravity''
has strongly adopted the geometric flavour of general relativity; it
speaks of surfaces and spaces, of geometries and shapes.

On the other hand, ``Lorentzian lattice quantum gravity'' has
irritated both brane theorists and general relativists (and more than
a few lattice physicists as well): It does not have, and does not seem
to require, the complicated superstructure of supersymmetry and all
the other technical machinery of brane theory/ string theory. (A
critically important feature of brane theory/ string theory which
justifies the amount of time spent on the model is that in an
appropriate limit it seems to approximate key aspects of general
relativity; and do so without the violent mathematical infinities
encountered in most other approaches. Of course, there is always the
risk that there might be other less complicated theories out there
that might do an equally good job in this regard.)  Additionally,
``Lorentzian lattice quantum gravity'' irritates some members of the
relativity community by not including {\emph{all}} possible
4-dimensional geometries: The key ingredient that makes this
Lorentzian approach different (and successful, at last in a
lower-dimensional setting) is that it to some extent enforces a
separation between the notions of space and time, so that space-time
is really taken as a product of ``space'' with ``time''. It then sums
over the resulting restricted set of (3+1)-dimensional geometries; not
over all 4-dimensional geometries (that being the traditional approach
of the so-called Euclidean lattice quantum gravity).

Technically, Lorentzian lattice quantum gravity restricts the sum over
4-dimensional geometries to cover only that subset of 4-dimensional
geometries compatible with the existence of 3+1 space+time dimensions.
(The condition used is a discretized version of stable causality; in
the sense of the existence of a global time function.)  The result of
this topological/ geometrical restriction is that the model produces
reasonably large, reasonably smooth patches of spacetime that look
like they are good precursors for our observable universe. (Euclidean
lattice quantum gravity, and variants thereof such as Matrix theory,
have an unfortunate tendency to curdle into long thin polymer-like
strands that look nothing like the more or less flat spacetime in our
immediate vicinity; Quantum geometry based on new variables likewise
encounters technical difficulties in generating an approximately
smooth manifold in the low-energy large-distance limit.)

The good news is that once reasonably large, reasonably flat, patches
of spacetime exist, the arguments leading to Sakharov's notion of
``induced gravity'' almost guarantee the generation of a cosmological
constant and an Einstein--Hilbert term in the effective action through
one-loop quantum effects [3]; and this would almost
automatically guarantee an inverse-square law at very low energies
(large distances). The bad news is that so far the large flat regions
have only been demonstrated to exist in 1+1 and 2+1 dimensions --- the
(3+1)-dimensional case continues to pose considerable technical
difficulties.

All in all, the development of ``Lorentzian lattice quantum gravity''
is extremely exciting: It is non-perturbative, definitely high-energy
(ultraviolet) finite, and has good prospects for an acceptable
low-energy (infra-red) limit. It has taken ideas from both the quantum
and the relativity camps, though it has not completely satisfied
either camp.  Keep an eye out for further developments.

{\bf References:}

Key papers on Lorentzian lattice quantum gravity:\\

[1]
J.~Ambjorn, A.~Dasgupta, J.~Jurkiewicz and R.~Loll,
``A Lorentzian cure for Euclidean troubles,''
Nucl.\ Phys.\ Proc.\ Suppl.\  {\bf 106} (2002) 977--979
\htmladdnormallink{{\tt arXiv:hep-th/0201104}}
{http://xxx.lanl.gov/abs/hep-th/0201104}

J.~Ambjorn, J.~Jurkiewicz and R.~Loll,
``3d Lorentzian, dynamically triangulated quantum gravity,''
Nucl.\ Phys.\ Proc.\ Suppl.\  {\bf 106} (2002) 980--982
\htmladdnormallink{{\tt arXiv:hep-lat/0201013}}
{http://xxx.lanl.gov/abs/hep-lat/0201013}.

J.~Ambjorn, J.~Jurkiewicz, R.~Loll and G.~Vernizzi,
``Lorentzian 3d gravity with wormholes via matrix models,''
JHEP 0109 (2001) 022 
\htmladdnormallink{{\tt arXiv:hep-th/0106082}}
{http://xxx.lanl.gov/abs/hep-th/0106082}

J.~Ambjorn, J.~Jurkiewicz and R.~Loll,
``Dynamically triangulating Lorentzian quantum gravity,''
Nucl. Phys. {\bf B610} (2001) 347--382 
\htmladdnormallink{{\tt arXiv:hep-th/0105267}}
{http://xxx.lanl.gov/abs/hep-th/0105267}.

A.~Dasgupta and R.~Loll,
``A proper-time cure for the conformal sickness in quantum gravity,''
Nucl.\ Phys.\ B {\bf 606} (2001) 357--379
\htmladdnormallink{{\tt arXiv:hep-th/0103186}}
{http://xxx.lanl.gov/abs/hep-th/0103186}.

J.~Ambjorn, J.~Jurkiewicz and R.~Loll,
``Non-perturbative 3d Lorentzian quantum gravity,''
Phys.\ Rev.\ D {\bf 64} (2001) 044011
\htmladdnormallink{{\tt arXiv:hep-th/0011276}}
{http://xxx.lanl.gov/abs/hep-th/0011276}.

R.~Loll,
``Discrete Lorentzian quantum gravity,''
Nucl.\ Phys.\ Proc.\ Suppl.\  {\bf 94} (2001) 96--107
\htmladdnormallink{{\tt arXiv:hep-th/0011194}}
{http://xxx.lanl.gov/abs/hep-th/0011194}.

J.~Ambjorn, J.~Jurkiewicz and R.~Loll,
``Computer simulations of 3d Lorentzian quantum gravity,''
Nucl.\ Phys.\ Proc.\ Suppl.\  {\bf 94} (2001) 689--692
\htmladdnormallink{{\tt arXiv:hep-lat/0011055}}
{http://xxx.lanl.gov/abs/hep-lat/0011055}.

J.~Ambjorn, J.~Jurkiewicz and R.~Loll,
``A non-perturbative Lorentzian path integral for gravity,''
Phys.\ Rev.\ Lett.\  {\bf 85} (2000) 924--927
\htmladdnormallink{{\tt arXiv:hep-th/0002050}}
{http://xxx.lanl.gov/abs/hep-th/0002050}.

[2]
A survey of brane theory and quantum geometry:\\
G. Horowitz, 
``Quantum Gravity at the Turn of the Millennium'',\\
MG9 --- Ninth Marcel Grossmann meeting, Rome, Jul 2000,\\
\htmladdnormallink{{\tt arXiv:gr-qc/0011089}}{
http://xxx.lanl.gov/abs/gr-qc/0011089}.

[3]
Sakharov's induced gravity:\\
A.D. Sakharov,
``Vacuum quantum fluctuations in curved space and the theory of gravitation'',
Sov. Phys. Dokl. {\bf 12} (1968) 1040-1041;
Dokl. Akad. Nauk Ser. Fiz. {\bf 177} (1967) 70-71.

\vfill
\pagebreak
\section*{\centerline {
Gravitational-wave phenomenology at PennState}}
\addtocontents{toc}{\protect\medskip}
\addtocontents{toc}{\bf Conference reports:}
\addtocontents{toc}{\protect\medskip}
\addcontentsline{toc}{subsubsection}{\it  
Gravitational-wave phenomenology at PennState, by Nils Andersson}
\begin{center}
Nils Andersson, University of Southampton, UK
\htmladdnormallink{N.Andersson@maths.soton.ac.uk}
{mailto:N.Andersson@maths.soton.ac.uk}
\end{center}

\noindent
During a few surprisingly warm days in early November 2001, 86 keen 
gravitational-wave scientists gathered at the State College Days Inn 
for the first annual Gravitational Wave (GW) phenomenology 
workshop.  The event was an excellent celebration of the 
recently funded Physics Frontier Center --- the talks were of extremely 
high standard, and there were numerous fruitful discussions. 

Talks were given on topics covering most 
areas relevant for gravitational-wave physics. In fact, I think the meeting 
is well described as a serious attempt at figuring out what 
``gravitational-wave phenomenology'' might actually mean... 
The Tuesday morning 
started off with a session on astrophysics. Peter Meszaros provided an 
update on gamma-ray bursts and the possible connection to GWs. 
Particularly interesting here is the fact that there no 
longer seems to be an ``energy crisis'': The energy released generally seems 
to be within one order of magnitude of $10^{51}$ ergs. Tony Mezzacappa
told us about the most recent simulations of core collapse 
supernovae. He easily
convinced us that this is a very hard problem, involving plenty of ``dirty'' 
and not very well understood physics. Although he was not very 
optimistic 
about being able to do fully relativistic calculations in the near future, 
he indicated significant progress on understanding 
the stabilizing influence of multidimensional 
radiation transport. 
Basically there now seems to be a consensus in the collapse 
community: Supernovae simply don't explode!    
 Joan Centrella rounded off the morning session with a nice talk 
covering the range of issues from source modeling to data analysis. 
She highlighted recent results that indicate that the dynamical bar-mode 
instability in rapidly spinning neutron stars may be much longer lived than 
was thought a couple of years ago. This could be very good news for
observations! 

The afternoon session was focussed on
key problems in relativity. Saul Teukolsky discussed whether 
numerical relativity was ``on the right track''. The question
was motivated by the fact that LIGO (and other detectors) are 
due to come online and there still are no ``accurate'' template
signals for  black-hole mergers. However, as Saul made clear, there
has been a lot of progress recently. In particular, our understanding
of the fundamental lack of stability of the ADM formalism has been 
much improved. The one issue of major importance that must ultimately 
be faced was also discussed. Namely, how to formulate 
``astrophysical'' initial data. This is a very difficult problem, 
which requires serious attention. Abhay Ashtekar discussed recent work 
on isolated black-hole horizons. He outlined an exciting scheme wherein
the properties of individual black holes may be evaluated in the vicinity 
of the horizon. If this idea could be implemented numerically, it could 
prove of tremendous use for black-hole excision etcetera.
During the following coffee break, the main topic of discussion
was Saul Teukolsky's slides: The general consensus was that they must 
have come out in the wash, and that he would be well advised to use
permanent ink next time...
The day ended with a talk by Eanna Flanagan on the radiation reaction 
problem. The main challenge still concerns the general binary orbital 
evolution in the Kerr spacetime. How are we supposed to deal with the
Carter constant? Kip Thorne commented that the issue is becoming crucial 
as LIGO is only 5 months away. But when he then asked what people in the 
community were doing about it Eanna was saved by the bell (as the firealarm
went off!). 

In the evening Kip gave one of his vintage public lectures on gravitational
waves. It was extremely well attended and clearly a very popular event. 

The Wednesday began with another astrophysics session. Vicky Kalogera
discussed the constraints that the several different binary pulsars
pose on the general stellar population, and how this relates to
observable GWs from inspiraling binaries and gamma-ray bursts.  She
pointed out that GW observations could challenge current stellar
evolution models, eg. by finding black holes with masses above
$15-20M_\odot$. Steinn Sigurdsson followed this with a discussion of
stochastic GW backgrounds, both primordial and astrophysical.  He
discussed the fact that GW has an ``Olber's paradox'' in that the
summed strain from all sources is not divergent. Recent estimates by
Sterl Phinney were discussed at length. These suggest that it is
because LIGO has difficulty seeing point sources that there can't be a
significant astrophysical background. If you see plenty of sources,
the background will swamp the detector. Given the number of galactic
stellar binaries this could provide a severe problem for LISA, and
people are now thinking hard about how accurately one can hope to
filter out the strongest binary signals from the LISA data stream in
order to unveil the primordial background. The morning session
finished by Alex Wolszczan describing how the radio technology is
improving towards the point where one should be able to detect GWs
from relativistic binaries.  This would require measurements of the
pulse arrival time to $\mu$s precision. Alex suggested that this might
a serious possibility on the 5 year timescale.

The afternoon was mysteriously labeled ``interface''. First, Joel
Tohline gave an overview of hydrodynamical simulations of various
relevant scenarios; close binary merger, bar-mode and r-mode
instabilities.  These simulations provide an impressive demonstration
of large scale numerics leading to new insights about the detailed
physics.  Various codes have now reached the level of reliability
where one can probe the truly nonlinear regime for quite realistic
scenarios.  I find that extremely exciting! Finally, Sam Finn gave the
last proper talk of the day. He discussed how GW observations could
provide tests of general relativity. As he put the question: ``Is
there any value added by testing the theory in the dynamical
sector?''.  Sam provided three cases that would provide very useful
information: Binary inspiral for inferring the mass of the graviton as
well as mapping the actual black-hole spacetime, and black-hole
ringdowns for providing unequivocal evidence of the presence of black
holes.  The day ended with a round-table discussion of GW
phenomenology.  The consensus seemed to be that we should take this to
mean ``the use of GW to explore astrophysics'', which makes a lot of
sense.

In the evening we were treated to a banquet at the Nittany Lion Inn.
It was a memorable event, with several entertaining speeches
describing Richard Isaacson's role in supporting gravity research (in
view of his retirement from NSF).

The talks on the final day concerned detector technology.  Massimo Cerdonio 
summarized that status of existing bar detectors and described possible future 
advances in technology. Most exciting here is the prospect for dual spheres, 
which would in principle provide acoustic broadband instruments with 
sensitivity in the kHz regime. Alessandra Buonnanno provided a peak into the 
future of advanced interferometers. A main issue for future generations 
concerns beating the standard quantum limit. Alessandra described this problem 
and discussed how one could hope to how to beat it. 
Robin Stebbins described the modeling plan for the LISA mission.
Appropriately he ended the talks of the meeting on a high note by pointing 
out that 75\% of all large scale NASA mission have failed. This is a thought 
too terrible to contemplate...

Kip Thorne closed the meeting with a succinct summary of the various
talks, thus putting everything in context. He concluded his overview
by wishing the new PFC luck in the future, and I would like to second
that here. The meeting was an exciting one, and the organizers deserve
a lot of credit. Not only did they provide a pleasant atmosphere, they
also left plenty of time for discussions in the busy program. This
should serve as a good example for organizers of future meetings: Get
people debating and you will have a great couple of days.

\bigskip
\bigskip

\begin{figure}[h]
\epsfxsize=4in
\epsfbox{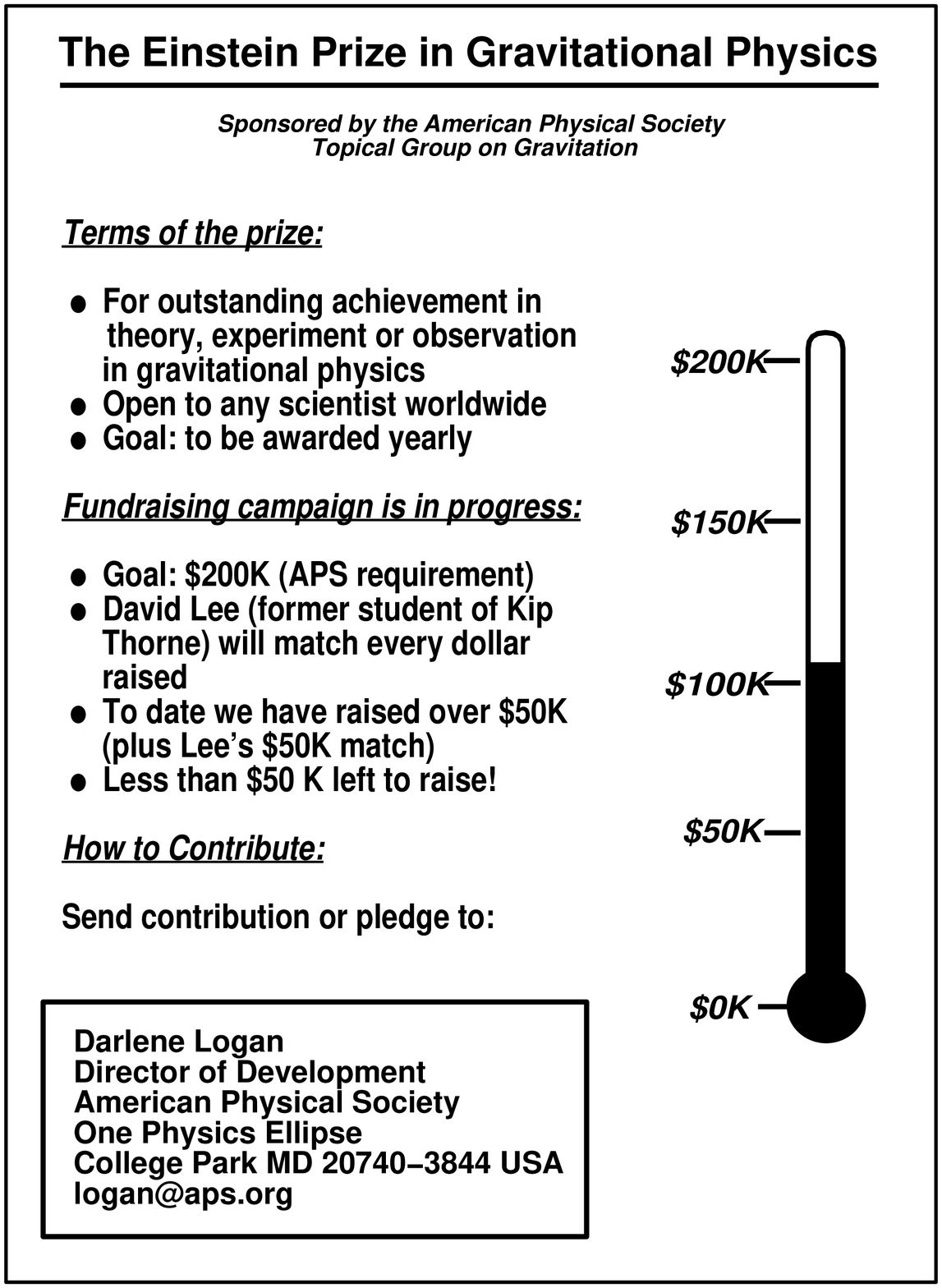}
\end{figure}

\end{document}